\documentclass[aps,amsfonts,nofootinbib,superscriptaddress,preprint]{revtex4-1}
\usepackage[normalem]{ulem} 
\makeatletter
\usepackage{amsmath}
\usepackage{amssymb}
\usepackage{amsbsy}
\usepackage{bm}
\usepackage{mathtools}
\usepackage{epsfig} 
\usepackage{color}
\usepackage{slashed}
\usepackage{soul}
\usepackage{comment}
\usepackage{appendix}
\usepackage{epsfig} 
\usepackage{hyperref}
\usepackage{cleveref}
\usepackage[T1]{fontenc}
\usepackage{caption}
\usepackage{subcaption}
\usepackage{graphicx}
\usepackage{multirow}

\newcommand{\stkout}[1]{\ifmmode\text{\sout{\ensuremath{#1}}}\else\sout{#1}\fi}
\newcommand{\bb}{\begin{equation}}
\newcommand{\ee}{\end{equation}}
\newcommand{\eqb}{\begin{eqnarray}}
\newcommand{\eqf}{\end{eqnarray}}

\def\veff{ {V_{\mbox{\tiny{eff}}}}}

\begin{document}
\title{  Resonance of Gravitational Axions-like Particles }

\author{Jorge Gamboa}
\email{jorge.gamboa@usach.cl}
\affiliation{Departamento de F\'{\i}sica, Universidad de Santiago de Chile, Santiago 9170020, Chile}
\author{Fernando Mendez}
\email{fernando.mendez@usach.cl}
\affiliation{Departamento de F\'{\i}sica, Universidad de Santiago de Chile, Santiago 9170020, Chile}

\begin{abstract}
The motion of gravitational axion-like particles (ALP) around a Kerr
black hole is analyzed, paying attention to resonance and distribution
of spectral radiation. We first discuss the computation of
$\sqrt{g}{\tilde R}_{\mu \nu \rho \rho \sigma}R^{\mu \nu \rho \sigma}$
and its implications with Pontryagin's theorem and a detailed analysis
of Teukolsky's master equation is done. After carefully analyzing the Teukolsky master equation, 
we show that this system exhibits resonance when $\omega \gtrsim \mu$ where $\mu$ 
is the mass of the ALP. {{A skew-normal distribution can approximate the energy distribution, and we can calculate the mean lifetime of the resonance for black holes with masses between 100 to 1000 $M\textsubscript{\(\odot\)}$. This range corresponds to a duration between $10^{-1}$s and $10^{41}$s, the observation range used in LIGO data.}}

\end{abstract}
\maketitle

\section{Introduction}

Dark matter permeates much of the current cosmology and particle
physics research because it can help solve many long-standing
problems. However, the search for dark matter encounters difficulties
along the way, and so far, one of the most plausible candidates is of
very light particles called axions.
 
 Axions are pseudoscalars that were proposed in
 \cite{peccei,weinberg,wilczek} to solve the strong CP problem and
 have become the cornerstone of modern particle physics and cosmology.
 
 The axion is described by 
 \bb
 {\cal L} \subset \frac{1}{2} (\partial \varphi)^2 -\frac{1}{2} m^2
 \varphi^2 +  g \varphi {\tilde F}_{\mu \nu}F^{\mu \nu} +
 \cdots, \label{a1}    
 \ee
 where ${\tilde F}_{\mu \nu}F^{\mu \nu}$ is the Pontryaguin density
 for electromagnetic field, $F_{\mu \nu}=\partial_\mu A_\nu -
 \partial_\nu A_\mu$ and $g$ is a coupling constant with dimension
 $-1$.
 
 The nature of the interaction $\varphi {\tilde F}_{\mu \nu}F^{\mu
   \nu}$ implies that $\varphi $ is a pseudoscalar and the solutions
 of the equation (plus Maxwell equations)
 \bb
 (\Box^2 +m^2)\varphi = g  \varphi  {\tilde F}_{\mu \nu}F^{\mu \nu}, \label{a2}
 \ee
 provide the ingredients for axion detection arguments \cite{sikivie}.  
   
  In this research, we would like to focus on a different coupling;
  namely, let us consider the replacement   
 \bb 
 F^{\mu \nu} {\tilde F}_{\mu \nu}  \to R^{\mu \nu \rho \sigma} {\tilde
   R}_{\mu \nu \rho \sigma}, 
 \label{1}
 \ee
 where $R^{\mu \nu \rho \sigma} {\tilde R}_{\mu \nu \rho \sigma}$ is
 the Pontryaguin-Riemann density, which is  
 \bb 
 R^{\mu \nu \rho \sigma} {\tilde R}_{\mu \nu \rho \sigma}=
 \epsilon^{\rho \sigma \alpha \beta} R_{\mu \nu \rho \sigma} R^{\mu
   \nu}_{\alpha \beta}.   
  \ee 
 
  This kind of system will obey the following system of equations
  \eqb 
  \left(\Box^2 + m^2 \right) \varphi &=& {\bar g} R^{\mu \nu \rho
    \sigma} {\tilde R}_{\mu \nu \rho \sigma}, \label{b1} 
  \\
  G^{\mu \nu} + C^{\mu \nu} &=& T^{\mu \nu}, \label{b2}
  \eqf 
 where $\Box^2= \frac{1}{\sqrt{-g}}\partial_\mu (g^{\mu
   \nu}\partial_\nu)$ is the Laplace-Beltrami operator,   ${\bar g}$
 is a coupling constant and $C^{\mu \nu}$ is  defined as \cite{jackiw}
 (Cotton's tensor) 
  \bb 
 C^{\mu \nu} =  \nabla_\alpha \varphi \, \epsilon^{\mu \beta \gamma
   (\mu} \nabla_\gamma R^{\nu)}_\beta + \left( \nabla_\sigma \nabla_\lambda\right) \varphi
 {\tilde R}^{\lambda (\mu \nu) \sigma},  
 \ee
 with $T_{\mu \nu}$ the energy-momentum tensor for the (pseudo)scalar
 field in a curved background. 
 
 At first sight, the system above retains many properties of the
 conventional axion but also differs substantially because when
 coupled to gravity, it becomes dynamically a very different system,
 These gravitational axions will be denoted generically as ALP.
 Additionally, the coupling (\ref{1}) is physically well-motivated
 \cite{witten} by the gravitational anomaly and, in analogy with the
 chiral anomaly where $\pi_0 \to 2 \gamma$ \cite{adler,jack1}, we
 might also expect the decay $\varphi \to 2 g$, where $g$ are
 gravitons. 
      
 In this paper, we will study the problem of ALP in a Kerr black-hole
 background, and we will focus mainly on resonance and radiation. 
 There are two reasons to consider carefully
 the phenomenon of resonance; the first is because our research is
 probably the first example in which Teukolsky's master equation can
 be explicitly worked out order by order, and resonance could be a
 manifest phenomenon; the second reason is that the careful analysis
 of the resonance allows not only to extract information about the
 properties of the ALPs but also --if the resonance occurs--
 it can be seen directly from the spectral radiation curves.

It's important to note that while Detweiler also examined Kerr's black holes in a different  context 
in \cite{detweiler}, his findings are not relevant to our current discussion due to various 
technical reasons that are unique to the Pontryaguin source we are utilizing.

The paper is structured as follows: in section II, we will begin by studying scalar perturbations and focus on the technical details of the problem. 
Section III will consider scalar perturbations and their implications with ALP. We will also explain the separation of variables of the Teukolsky equation. 
Section IV will explain the radial equation with a source in detail and solve it asymptotically. 
In section V, we will study the emission of gravitational radiation by axion-like particles and numerically calculate the spectral distribution of radiation. 
Finally, in section VI, we will give our discussions and conclusions. The $S_{\ell}(x)$ properties and essential formulas are in an appendix.  

\section{Axions as scalar perturbations}
  
  In this section, we address the problem of solving the equation for
  axion-like particles in a Kerr background with a Pontryaguin
  source. The no-source case has been discussed for a long time by Press
  and Teukolsky \cite{press} and Damour et al \cite{damour}, Dolan in \cite{dolan}, and
  an updated reference can be found in \cite{myung,yamada}.
  
  However, Detweiler in \cite{det3} developed a calculation strategy
  that seems to us to better fit our purpose and that we will use
  here.  Basically, the idea developed in \cite{detweiler, det3} is to
  consider a Klein-Gordon equation in a Kerr background, and instead
  of looking for exact solutions, asymptotic solutions can be analyzed
  to capture the essential physical aspects.

 The action is 
 \begin{equation}
 S= \int d^4 x \sqrt{-g} \left[ R +  \bar{g}\, R^{\mu \nu \rho \sigma}
   {\tilde R}_{\mu \nu \rho \sigma} +  (\partial \varphi)^2  + \cdots\right].  
 \label{act1}
 \end{equation}
with $\bar{g}$ the coupling constant. The Kerr metric is assumed,  and
in Boyer-Lindquist coordinates, it is 
 \eqb
 ds^2 &=&   -\left(1-\frac{2Mr}{\rho^2} \right) dt^2 -\frac{4 M a\,
   r}{\rho^2} \sin^2 \theta d\phi dt  \nonumber  
 \\
 &+& \left(r^2 +a^2 + \frac{2M r a^2\sin^ 2 \theta}{\rho^2}\right)
 \sin^ 2 \theta\, d\phi^2 +  \frac{\rho^2 }{\Delta} dr^2 
 + \rho^2 \,d \theta^2,    \label{b3}
 \eqf
 where 
 \eqb 
 \rho^2 &=& r^2 +  a^2 \cos^2 \theta, \nonumber
 \\
  \Delta &=& r^2 -2 M r + a^2, \label{met1}
 \eqf 
 and $a= \frac{J}{M}$ relates the angular momentum with the mass of
 the black hole.  
 
 Note that when $a\to 0$, the angular momentum vanishes, and the metric
 (\ref{met1}) reduces to the Schwarschild one.  The singularities appear when 
 $\Delta =0$, and we have the event horizons  
 \bb 
 r_{\pm} = M\pm \sqrt{M^2 -a^2}, \label{met2}
 \ee 
 which correspond to the inner and outer event horizons. 
 
 The relation $\rho=0$ implies that for $r\to 0$ and $\theta \to
 \frac{\pi}{2}$, the metric component $g_{tt}\to \infty$. is the true
 singularity of the Kerr metric. Indeed, the Kretschmann scalar $ K =
 R_{\alpha\beta\gamma\mu} R^{\alpha\beta\gamma\mu}$ for $r \to 0$ is
 $K_{r\to 0} \propto M^2\,\sec^6\theta $, showing that
 $\theta\to\pi/2$ (together with $r\to0$) is a singularity independent
 of coordinates.

The calculation of the source term for the scalar field given  the action (\ref{act1}), 
in the Kerr background, yields \cite{yunes}
 \bb
  \sqrt{-g} R^{\mu \rho \sigma} {\tilde R}_{\mu \nu \rho \sigma} =
  -96 M^2 a\,\frac{ r \cos \theta \sin \theta}{(r^2 +  a^2 \cos^2
    \theta)^5} (3 r^2 - a^2 \cos^2  \theta)(r^2 -3 a^2 \cos^2 \theta). 
  \label{p1}
 \ee
 
\subsection{Pontryagin theorem and subtleties }
 Equation (\ref{p1}), although correct, cannot be complete because,
 otherwise, the topological properties of a Kerr black hole would have
 no physical effect. Several reasons indicate that this is not the
 case and vorticity is an example that indicates that a turbulent
 stage of a Kerr black hole must be important in the final dynamics.

 Although we will not address the turbulence problem, we would like to
 point out that the analog  of quantized circulation is   
 \bb
 \int d^4 x ~ \sqrt{-g} ~R^{\mu \rho \sigma} {\tilde R}_{\mu \nu \rho \sigma} = n, \label{pp1}
 \ee
 where $n=0,\pm 1,\pm 2, \cdots$ and (\ref{pp1})  is a standard theorem in geometry \cite{eguchi}. 
 
 In our case, the direct calculation yields
  \bb
 \int d^3 x ~ \sqrt{-g}~ R^{\mu \rho \sigma} {\tilde R}_{\mu \nu \rho \sigma} = 0, \label{pp11}
 \ee 
since due to (\ref{p1}) the Pontryagin density depends only on $r$ and $\theta$. 

The first feeling is that a static metric (stationary in this case)
does not induce topological properties and, therefore (\ref{pp11}) vanishes and the winding number $n=0$.
  
 However, if $n\neq 0$, the integral (\ref{pp11}) is not well defined
 for a stationary metric, and we should regularize it using some
 reasonableness criterion.  Which criterion?, We think it is enough
 that Pontryagin's theorem is satisfied.  
 
Thus, we propose the following modification for the Kerr metric
 \bb 
 \sqrt{-g}~ R^{\mu \rho \sigma} {\tilde R}_{\mu \nu \rho \sigma} \to
 \sqrt{-g}~ R^{\mu \rho \sigma} {\tilde R}_{\mu \nu \rho \sigma}
 \delta (x_0),  
  \ee
 which is consistent with  (\ref{pp1}).
 
 The above result has a very interesting physical implication because
 the factor  $\delta (t)$ correctly defines the integral on the
 four-manifold and  induces an initial condition to produce
 gravitational radiation.  
 
 Two technical aspects are responsible for these consequences, namely.
 i) since the source depends on $r$ and $\theta$, the angular momentum
 along $\varphi$ is conserved, and the general solution of the
 Teukolsky master equation is a function of $r,\theta$ and $t$; ii)
 since the LHS is time-dependent, the presence of the
 $\delta$-function in the RHS becomes mandatory. 
  
 \section{Scalar Perturbations}

 After discussing these mathematical issues, scalar perturbations for
 a Kerr black hole can all be written in terms of the Teukolsky master
 equation \cite{teukolsky}, which,  for the scalar case, reads 
 \begin{eqnarray}
&& \frac{\partial}{\partial r} \left(\Delta \frac{\partial
     \Phi}{\partial r}\right) -\frac{a^2}{\Delta} \frac{\partial^2
     \Phi}{\partial \varphi^2} -\frac{4 M r a}{\Delta}
   \frac{\partial^2 \Phi}{\partial \varphi \partial t} -
   \left(\frac{(r^2 +a^2)^2}{\Delta} -a^2 \sin^2 \theta \right)
   \frac{\partial^2 \Phi}{\partial t^2}  
\nonumber
\\ 
&& +  \frac{1}{\sin \theta} \frac{\partial}{\partial \theta} \left(
\sin \theta \frac{\partial \Phi}{\partial \theta}\right) +
\frac{1}{\sin^2 \theta} \frac{\partial^2 \Phi}{\partial \varphi^2} - 
\mu^2 (r^2+ a^2 \cos^2 \theta) \,\Phi =  
\kappa\,(r^2+ a^2 \cos^2 \theta)\,{\tilde R}_{\mu \nu \rho
  \sigma}R^{\mu \nu \rho \sigma}\,\delta(t),  
\nonumber
\\
&&\equiv{T}(x)
\label{apteukolsky}
\end{eqnarray} 
with $\kappa$ a constant with canonical dimension $-2$ so that LHS and
RHS of the previous equation has dimension ${+1}$.  

The source term ${T}(x)$ turns out to be
\begin{eqnarray}
T(x) &=&
 96\kappa M^2\,a\frac{ r \cos \theta }{(r^2 + a^2 \cos^2 \theta)^5}
 (3 r^2 - a^2 \cos^2 \theta)(r^2 -3 a^2 \cos^2 \theta)\,\delta(t),
 \nonumber
 \\
 &\equiv&{\cal T}(r,\theta)\,\delta(t).
\label{app1}
\end{eqnarray}

Since the source is $\varphi$-independent, we look for solutions
$\Phi(t,r,\theta)$ so that equation  
(\ref{apteukolsky}) reads
\begin{eqnarray}
\frac{\partial}{\partial r} \left(\Delta \frac{\partial \Phi}{\partial
  r}\right)-\left(\frac{(r^2 +a^2)^2}{\Delta} -a^2 \sin^2 \theta
\right) \frac{\partial^2 \Phi}{\partial t^2} - \mu^2 (r^2+ a^2 \cos^2
\theta)\Phi & &  
\nonumber
\\ 
+  \frac{1}{\sin \theta} \frac{\partial}{\partial \theta} \left( \sin
\theta \frac{\partial \Phi}{\partial \theta}\right)  &=& 
{\cal T}(r,\theta)\,\delta(t).
\nonumber
\label{apteukolsky2}
\end{eqnarray} 

Then, we look for solutions with the form
\begin{equation}
\label{expan1}
\Phi(t,r,\theta) = \frac1{2\pi}\sum_\ell \int e^{\imath{{\omega}} t}
R_\ell (r)S_\ell(c,\theta)\,d\omega,  
\end{equation}
where the angular function $S_\ell(c,\theta)$ satisfies the equation
\cite{abra,mike,lammer2014spheroidal} 
\begin{equation}
\label{oblate}
\frac{1}{\sin \theta}\frac{d}{d \theta} \left[ \sin\theta
  \frac{S_{\ell}}{d\theta}\right] + \left(\lambda_{\ell} +c^2  \cos^2
\theta\right) S_{\ell} =0, 
\end{equation}
with $c^2= a^2(\omega^2 -\mu^2)$,  
and $\lambda_\ell$ is  
the  separation constant, which must be determined (see Appendix
\ref{sec:appendix1} for  details).   

The  radial equation reads
\begin{equation}
	\label{appradial}
	\frac{d}{dr} \left(\Delta \frac{d R_\ell}{d r}\right)+\left[ 
	\frac{\omega^2(r^2 +a^2)^2}{\Delta} - ( \mu^2 r^2 +\omega^2 a^2 +
	\lambda_{\ell}) \right]R_{\ell} = A_\ell(r),  
\end{equation}
with $A_\ell$ defined through
\begin{equation}
	A_\ell(r) = \int \,{\cal T}(r,\theta)\,S^*_\ell(c,\theta)\,d(\cos\theta),
\end{equation}
that is, the coefficients of the source, $\cal T$,  spanned in the base $S_\ell$.

Explicitly,
\begin{equation}
A_\ell(r) =96\,\kappa\, M^2\,a\,\,r\,\int_{-1}^{1}\frac{
  x(r^2 -3 a^2\,x^2) }{(r^2 + a^2 \, x^2)^5}(3 r^2 - a^2
\,x^2)\,{S^*}_\ell( c,x) dx,
\label{exact} 
\end{equation}

It is hard to find analytical solutions of (\ref{appradial}), so let's do some proper redefinitions. It is convenient to define 
dimensionless variables.
$$
y= \frac{r}{M},\quad \delta = \frac{a}M,\quad {\bar{\omega}}=\omega\,M,\quad
{\bar \mu} =\mu\,M,
$$
so that the radial equation reads now
\begin{equation}
\label{appradial2}
\frac{d}{dy} \left(\Delta(y) \frac{d R_\ell}{d y}\right)+\left[ 
{\bar{\omega}}^2\frac{(y^2 +\delta^2)^2}{\Delta(y)} - ( {\bar \mu} ^2\, y^2 +
{\bar{\omega}}^2\delta^2 + \lambda_{\ell}) \right]R_{\ell} = A_\ell(y),  
\end{equation}
with
$$
\Delta(y) = y^2 - 2 y + \delta^2, 
$$
and
$$
A_\ell(y) = 96\,\frac{\kappa}{ M^2}\,y\,\delta\, 
\int_{-1}^{1}\frac{  x(y^2 -3 \delta^2\,x^2) }{(y^2 + \delta^2 \,
  x^2)^5}(3 y^2 - \delta^2 \,x^2)\,{S^*}_\ell(c,x) dx. 
$$
Note that  
\bb 
c^2= a^2(\omega^2-\mu^2) = \delta^2({\bar{\omega}}^2-{\bar \mu}
^2). \label{cdef} 
\ee 

Finally, note also that here
$M$ has dimensions of energy$^{-1}$ and therefore, $A_\ell$ has
dimensions of energy, the same dimension as $R_\ell(y)$.
Then we define
\begin{equation}
Y_\ell(y) = \frac{M^2}{96 \kappa}\,R_\ell(M y),
\end{equation}
and then, the  fully dimensionless radial equation can be
written as  
\begin{eqnarray} 
\Delta\frac{d}{dy} \left(\Delta \frac{d Y_\ell}{d y}\right)&+&
\left[ 
{\bar{\omega}}^2(y^2 +\delta^2)^2 - \Delta( {\bar \mu}^2\, y^2 +
{\bar{\omega}}^2\delta^2 + \lambda_{\ell}) \right]Y_{\ell} 
\nonumber
\\
&=&\frac{M^2}{96\kappa}\Delta\,A_\ell
\nonumber
\\
&=& y\,\delta\,\Delta\, 
\int_{-1}^{1}\frac{  x(y^2 -3 \delta^2\,x^2) }{(y^2 + \delta^2 \,
  x^2)^5}(3 y^2 - \delta^2 \,x^2)\,{S^*}_\ell( c,x) dx. 
\nonumber
\\
&\equiv & \tilde{A}_\ell(y).
\label{appradial3}
\end{eqnarray}
  
The following sections are devoted to the study of numerical solutions to this 
equation, and also to the analysis of asymptotic structure.

\section{Radial Equation: Asymptotic analysis}
In this section, we will make an asymptotic analysis of the radial
equation, which, as we show below, has important physical consequences
in the Teukolsky master equation for pseudoscalar
fields.

In effect, Pontryagin's term is a very special source because being
in the case that we consider a function of the form $F(r,\theta)$, it
implies that for even values of $\ell$ the source vanishes while for odd
values, this is not the case.

    To analyze the asymptotic regions, we first change coordinates to
    {\it tortoise} coordinates $y_*$ defined through
    \bb
    \frac{dy_*}{d y} = \frac{y^2 +\delta^2}{\Delta(y)}=\frac{y^2 +\delta^2}{y^2-2 y + \delta^2}. 
    \label{redef}
    \ee
    Equation  (\ref{appradial3})  reads 
    \begin{equation}
    	\label{radialtortoise}
    	(y^2+\delta^2)\frac{d}{dy_*} \left((y^2+\delta^2) \frac{d Y_\ell}{d y_*}\right)+
    	\left[ 
    	{\bar{\omega}}^2(y^2 +\delta^2)^2 - \Delta( {\bar \mu}^2\, y^2 +
    	{\bar{\omega}}^2\delta^2 + \lambda_{\ell}) \right]Y_{\ell}=\tilde{A}_\ell(y),
    \end{equation}
    where all functions of $y$ are understood as functions of $y_*$  trough $y= y(y_*)$,
    while $Y_\ell(y_*) \equiv R(y(y_*))$. By defining the function $W(y_*)$ through
    $$
    Y_\ell(y_*)= (y^2+\delta^2)^{-1/2}\,W_\ell(y_*),
    $$   
    equation (\ref{radialtortoise}) is
    \begin{equation}
    	\frac{d^2\,W_\ell}{dy_*^2}+
    	\left[
    	{\bar{\omega}}^2 - \frac{\Delta}{(y^2+\delta^2)^2}
    	\left( 
    	{\bar \mu}^2\, y^2 + {\bar{\omega}}^2\delta^2 + \lambda_{\ell}  +
    	\frac{1}{(y^2+\delta^2)^2}\left(\Delta\,\delta^2+2y(y^2-\delta^2)
    	\right)\right)
    	\right]
    	\,W_\ell=\bar{A}_\ell(y_*),
    	\label{rede12app} 
    \end{equation}
    with
    \begin{equation}
      \label{sourtortl}
	\bar{A}_\ell(y_*)=\frac{\tilde{A}_\ell(y)}{(y^2+\delta^2)^{3/2}}.
    \end{equation}

    That is, $W_\ell$ satisfies the  Schr\"odinger-type equation
    \begin{equation}
    	\label{schtype}
    	\frac{d^2\,W_\ell}{dy_*^2} + \veff(y_*)
    	\,W_\ell=\bar{A}_\ell(y_*).
    \end{equation}
        We are interested in the solutions of (\ref{schtype})   in the
    asymptotic regions  $y_*\to\infty$ (solution at infinity) and
    $y_*\to-\infty$ (the near horizon solution).
    
    We first study the behavior of the source in these limits.

    \subsection{The source}

    Since $S_\ell$ is a superposition of  Legendre's
    polynomial  $P_\ell(x)$  (see Appendix \ref{sec:appendix1}), the source 
    in (\ref{appradial3}) is 
    \begin{eqnarray}
      \tilde{A}_\ell &=& y\, \Delta\,\delta\, \int_{-1}^{1}\frac{  x(y^2 -3
        \delta^2\,x^2) }{(y^2 + \delta^2 \,        x^2)^5}(3 y^2 -
      \delta^2 \,x^2)\,{S^*}_\ell( c,x) dx, 
      \\
      \nonumber
      &=&y\,\Delta\,\delta\,\sum_{\ell'}B_{\ell,\ell'}(c) \int_{-1}^{1}\frac{  x(y^2 -3
        \delta^2\,x^2) }{(y^2 + \delta^2 \,          x^2)^5}(3 y^2 -
      \delta^2 \,x^2)\,P_\ell'(x) dx,
      \nonumber
      \\
      &=& y\, \Delta\,\delta\,\sum_{\ell'= 0}B_{\ell,2\ell'+1}(c) \int_{-1}^{1}\frac{  x(y^2 -3
        \delta^2\,x^2) }{(y^2 + \delta^2 \,          x^2 )^5}(3 y^2 -
      \delta^2 \,x^2)\,P_{2\ell'+1}(x) dx,
      \nonumber
      \\
      &=&\sum_{\ell'= 0}B_{\ell,2\ell'+1}(c)\, {\cal I}_{2\ell'+1}(y,\delta),
    \end{eqnarray}
    with
    \begin{equation}
      \label{If}
      {\cal I}_{2\ell'+1}(y,\delta) = y\, \Delta\,\delta\,\int_{-1}^{1}\frac{  x(y^2 -3
        \delta^2\,x^2) }{(y^2 + \delta^2 \,          x^2 )^5}(3 y^2 -
      \delta^2 \,x^2)\,P_{2\ell'+1}(x) dx.
    \end{equation}

    It can be shown that coefficients $B$ have the following property
    \begin{equation}
      B_{2m,2n+1} = 0 = B_{2n+1,2m}, \quad m,n \in \{0,1,2,\cdots\},
    \end{equation}
    and, therefore
    \begin{equation}
	\bar{A}_{2n} =0, \quad n \in \{0,1,2,\cdots\}
    \end{equation}
    
\begin{figure}[h!]
	\centering
	\begin{subfigure}[h!]{0.45\textwidth}
		\centering
		\includegraphics[width=\linewidth]{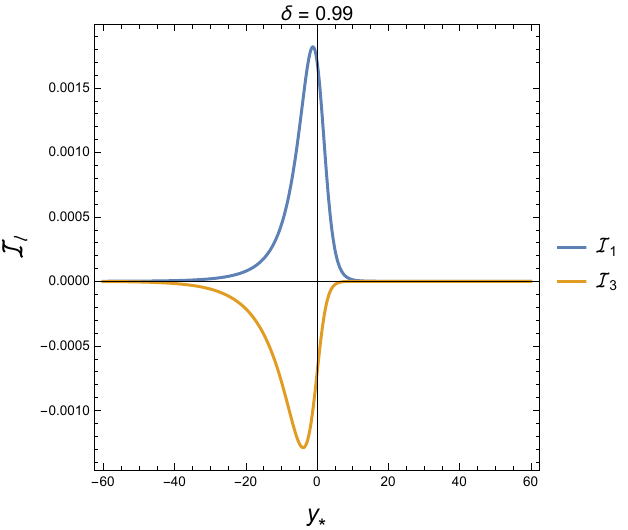} 
		\caption{Function ${\cal I}_\ell$ vs  rescaled tortoise
			coordinate $y_*$ for $\delta =0.99$ }
		\label{fig:1sub1}
	\end{subfigure}
	\hfill
	\begin{subfigure}[h!]{0.45\textwidth}
		\centering
		\includegraphics[width=\linewidth]{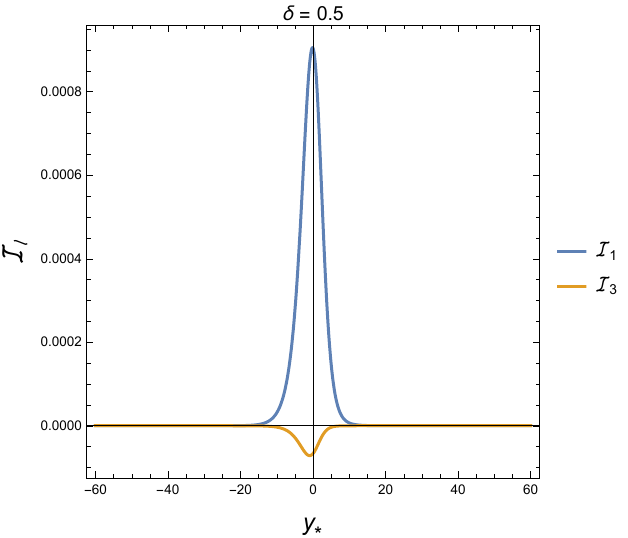}
		\caption{Function ${\cal I}_\ell$ vs  rescaled tortoise
			coordinate $y_*$ for $\delta =0.5$ }
		\label{fig:1sub2}
	\end{subfigure}
	\begin{subfigure}[h!]{0.6\textwidth}
		\centering
		\includegraphics[width=\linewidth]{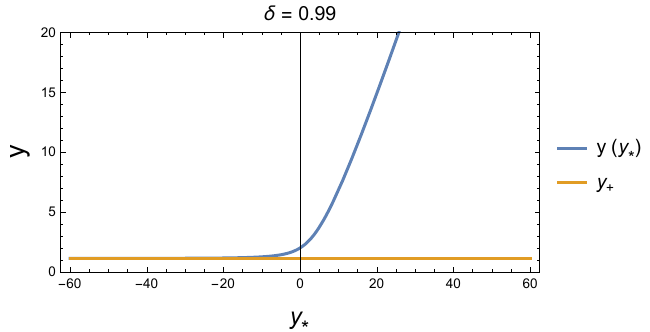}
		\caption{The coordinate $y$ as function of $y_*$ and the
			horizon $y_+$. }\label{fig:1sub3}
	\end{subfigure}
	\caption{Panels (a) and (b) show the function  ${\cal I}_\ell$ defined in (\ref{If}) for
		two different black hole's rotation velocity $\delta = a/M$
		for $\ell =1$, and $\ell=3$. Panel (c) shows $y(y_*)$ and the coincidence of $y_+$ (the horizon) 
		with $y_*\to -\infty$}  \label{fig:FIG1}
    \end{figure}

    In our numerical analysis, we consider $\ell =0,1,2,3$, then the
    relevant functions for us are ${\cal I}_1$ and ${\cal I}_3$, shown in
    Figure \ref{fig:FIG1} for two different values of $\delta$. The
    maximal contribution occurs in the region $y_* \lesssim 0 $, that is,
    toward the outer horizon, while contributions from $y_*\to\infty$ are
    negligible.

    In (\ref{fig:1sub3}) we can check that for $\delta =0.99$, the  horizon is
    reached at $y_*\approx 20$. Then, numerically, $|y_*| > 30$ is a good
    approximation for the limits $y_*\to\pm\infty$.

    The source terms in (\ref{schtype}) is, therefore, zero for even values of $\ell$, 
    while for the two other cases under analysis, they are
    \begin{eqnarray}
      \bar{A}_1(y_*) &=&(y^2+\delta^2)^{-3/2}(B_{1,1}(c)\, {\cal I}_1+B_{1,3}(c) \,{\cal I}_3),
      \\
      \bar{A}_3 (y_*)&=&(y^2+\delta^2)^{-3/2}(B_{3,1}(c)\, {\cal I}_1+B_{3,3} (c)\,{\cal I}_3),
    \end{eqnarray}
    with $B_{1,1},B_{1,3},B_{3,1},B_{3,3}$ given in Appendix \ref{sec:appendix1}, and 
    \begin{eqnarray}
      {\cal I}_1 &=&\frac{2y\delta\Delta(y^2-\delta^2)}{(y^2+\delta^2)^4},
      \\
        {\cal I}_3 &=&\frac{5\Delta}{2\delta^4} \tan ^{-1}\left(\frac{\delta }{y}\right)
        - \frac{y\Delta}{6\delta^3 \left(y^2+\delta ^2\right)^4}
        \left(15 y^6+55 y^4 \delta ^2+73 y^2 \delta ^4 + 57 \delta
        ^6\right).  
   \end{eqnarray}

\subsection{Numerical Solutions}

The potential $\veff$ in  the limits previously discussed has the
following asymptotic behavior  
    \begin{equation}
    	\label{asympV}
    	\veff=
    	\begin{cases}
    		k^2 +  {\cal O} (y^{-1}),   & y_* \to\infty ~~(y\to\infty),
    		\\
    		{\bar{\omega}}^2 +  {\cal O} (y-y_+), & y_*\to -\infty ~~(y\to y_+),
    	\end{cases}
    \end{equation}
with $k^2 = {\bar{\omega}}^2-\bar\mu^2\geq 0$. We can treat the equation as an
homogeneous equation in these limits  since the source can be taken
zero there, as shown in Fig. \ref{fig:FIG1}.

The asymptotic solutions are, therefore
\begin{eqnarray}
  W^{(+)} (y_*) &\sim& A^{(+)} \, e^{\imath {\bar{\omega}} y_*} + C^{(+)} \,
  e^{-\imath   {\bar{\omega}} y_*}
  \quad (y_*\to-\infty),
  \\
  W^{(\infty)}(y_*) &\sim& A^{(\infty)} \, e^{\imath k  y_*}
  + C^{(\infty)} \, e^{-\imath   k y_*}
  \quad (y_*\to\infty).
\end{eqnarray}

\begin{figure}[h!]
	\centering
	\begin{subfigure}[h!]{0.45\textwidth}
		\centering
		\includegraphics[width=\linewidth]{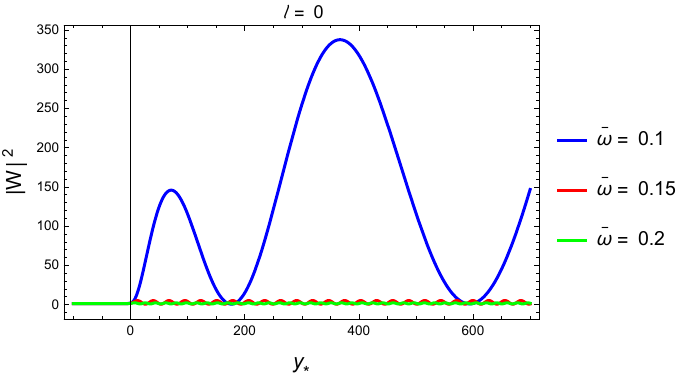}
		\caption{ $|W|^2$ for ${\bar{\omega}} \sim \bar\mu$ and $\ell=0$. }
		\label{fig:2sub1}
	\end{subfigure}
	\begin{subfigure}[h!]{0.45\textwidth}
		\centering
		\includegraphics[width=\linewidth]{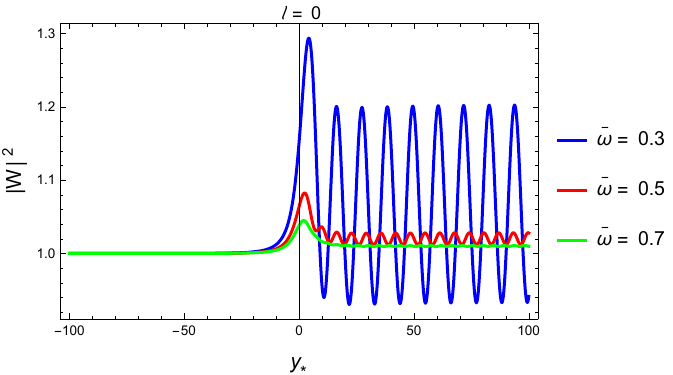}
		\caption{ $|W|^2$  for ${\bar{\omega}} > \bar\mu$ and $\ell=0. $ }\label{fig:2sub2}
	\end{subfigure}
	\begin{subfigure}[h!]{0.45\textwidth}
	\centering
	\includegraphics[width=\linewidth]{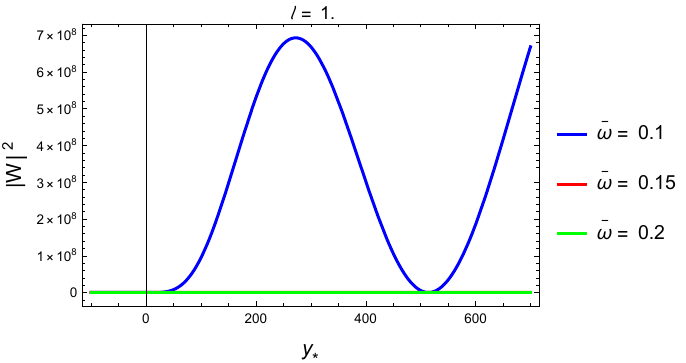}
	\caption{ $|W|^2$for ${\bar{\omega}} \sim \bar\mu$  and $\ell=1$.}\label{fig:2sub3}
\end{subfigure}
	\begin{subfigure}[h!]{0.45\textwidth}
	\centering
	\includegraphics[width=\linewidth]{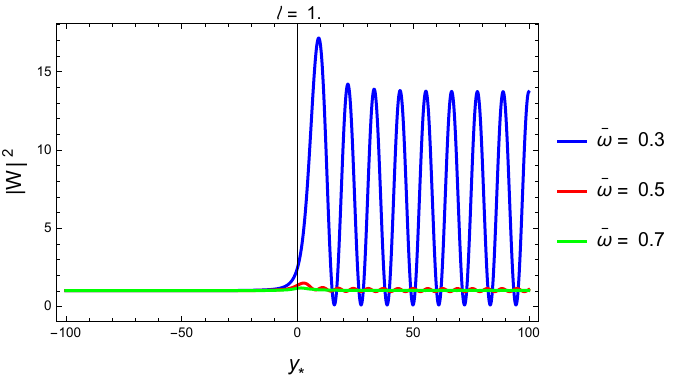}
	\caption{$|W|^2$ for ${\bar{\omega}} > \bar\mu$ and $\ell=1$. }\label{fig:2sub4}
\end{subfigure}
	\begin{subfigure}[h!]{0.45\textwidth}
	\centering
	\includegraphics[width=\linewidth]{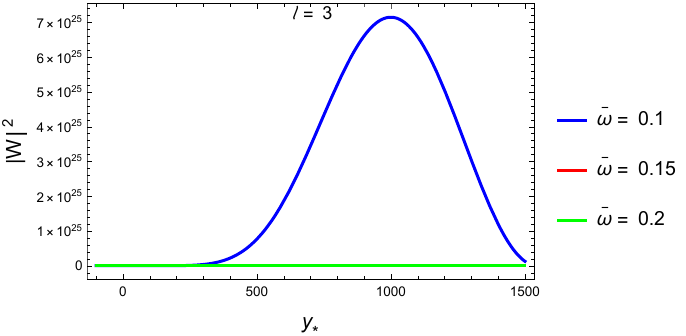}
	\caption{$|W|^2$  for ${\bar{\omega}} \sim  \bar\mu$, and $\ell=3$. }\label{fig:2sub5}
\end{subfigure}
	\begin{subfigure}[h!]{0.45\textwidth}
	\centering
	\includegraphics[width=\linewidth]{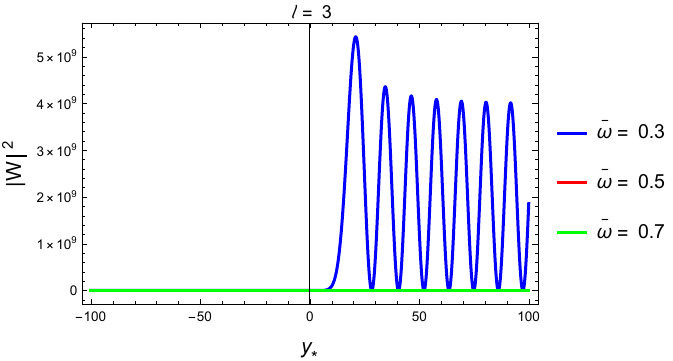}
	\caption{$|W|^2$  for ${\bar{\omega}} > \mu$, and $\ell=3$. }\label{fig:2sub6}
\end{subfigure}
	\caption{The amplitude $|W|^2$ for $\delta =  0.99$,  $\bar\mu=0.1$ and $\ell =0,1$ for different values of ${\bar{\omega}}$. }  \label{fig:FIG2}
\end{figure}

Following (\cite{detweiler}), we choose $A^{(+)}=1,C^{(+)} = 0$.
Numerical solutions for this choice are displayed in Fig (\ref{fig:FIG2}) for   $\ell =0,1$

As we pointed out before, the source components $\bar{A}_\ell$
are non-zero for $\ell$ odd (shown in Figures \ref{fig:1sub1} 
and \ref{fig:1sub2}). It is interesting to compare with the 
sourceless case. 

\begin{figure}[h!]
	\centering
	\begin{subfigure}[h!]{0.45\textwidth}
		\centering
		\includegraphics[width=\linewidth]{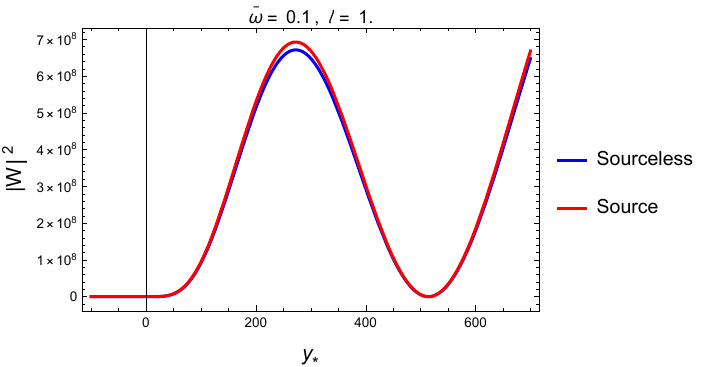}
		\caption{ $|W|^2$ for ${\bar{\omega}} \sim \bar\mu$ and $\ell=1$. Effect
		of the source term. }
		\label{fig:3sub1}
	\end{subfigure}
	\begin{subfigure}[h!]{0.45\textwidth}
		\centering
		\includegraphics[width=\linewidth]{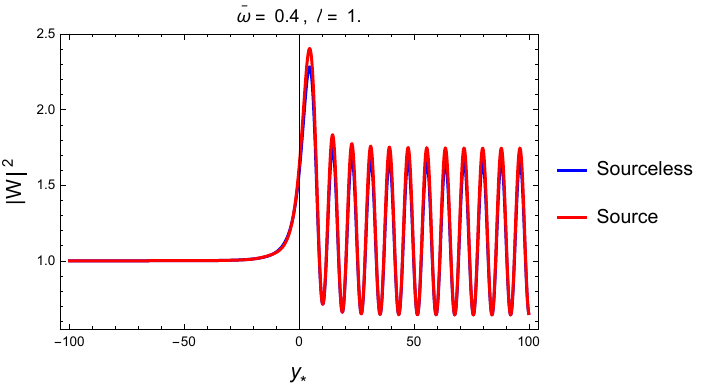}
		\caption{ $|W|^2$  for ${\bar{\omega}} > \bar\mu$ and $\ell=1$. Effect of the source term.}\label{fig:3sub2}
	\end{subfigure}
	\begin{subfigure}[h!]{0.45\textwidth}
	\centering
	\includegraphics[width=\linewidth]{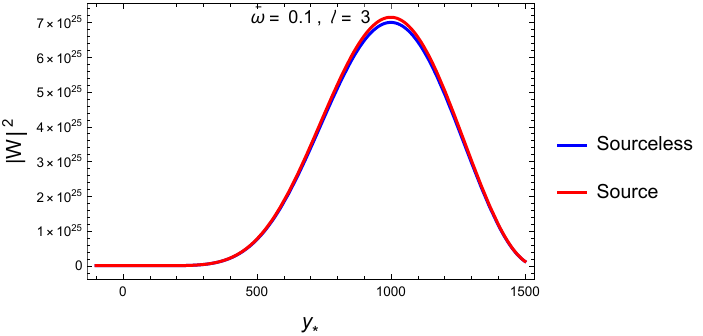}
	\caption{ $|W|^2$  for ${\bar{\omega}} \sim \bar \mu$ and $\ell=3$. Effect of the source term.}\label{fig:3sub3}
\end{subfigure}
	\begin{subfigure}[h!]{0.45\textwidth}
	\centering
	\includegraphics[width=\linewidth]{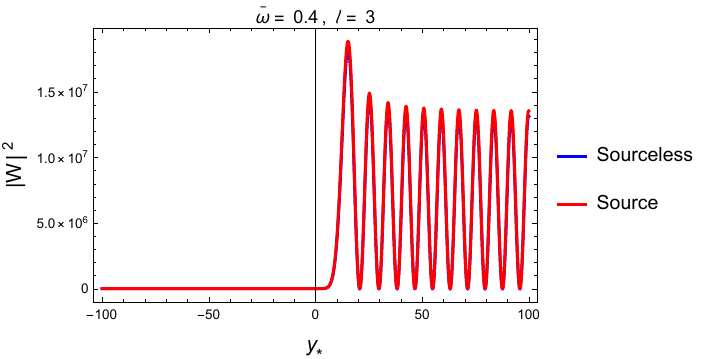}
	\caption{ $|W|^2$  for ${\bar{\omega}} > \bar \mu$ and $\ell=3$. Effect of the source term.}\label{fig:3sub4}
\end{subfigure}
	\caption{The amplitude $|W|^2$ for $\delta =  0.99$,  $\bar\mu=0.1$ and $\ell = 1, 3$ for the cases with and without source term }  \label{fig:FIG3}
\end{figure}

In Figure \ref{fig:FIG3}, we plot the case $\ell=1$ and $\ell=3$
for different values of ${\bar{\omega}}$, comparing the solution with and without source.

The source mainly affects the maxima (peaks) of $|W^2|$, but not 
the position of these peaks. Besides, the amplitude increases for
higher values of $\ell$, and the highest amplitudes occur for 
${\bar{\omega}}\sim \bar\mu$. This last condition, $c\sim 0$, corresponds to 
 the {\it longwave approximation}.
  
Indeed, the equation (\ref{exact})  is analogous to the partial wave method 
in quantum mechanics theory but spheroidal harmonics instead of Legendre polynomials.  

In (\ref{cdef}) we can write $c^2=a^2(\omega^2 -{\mu} ^2) = a^2|{p}|^2\sim \left(\frac{a}{\lambda}\right)^2$, where $|{p}|$ is the momentum of   the scalar field, and therefore  the limit $c\to 0$ is equivalent to
$a\ll \lambda$, which is the well-known long-wave approximation (LWA)
introduced by Isaacson \cite{isaacson} in gravitational
radiation \footnote{However, we emphasize, and we must not lose sight
of this, that $c\to 0$ must be understood, of course as $\omega
\approx \mu$.}.    
  
In this approximation
  \bb
  S_\ell(c,x) \approx P_\ell(x),  
  \ee
 and 
$\lambda_\ell\approx\ell(\ell+1)$. 

\begin{figure}[h!]
	\centering
	\begin{subfigure}[h!]{0.45\textwidth}
		\centering
		\includegraphics[width=\linewidth]{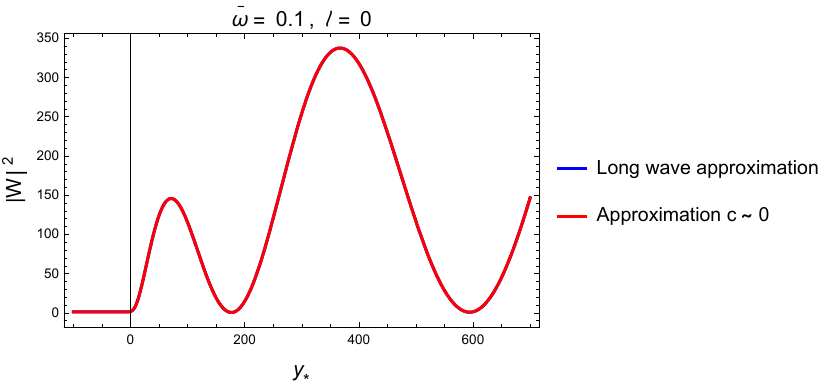}
		\caption{ $|W|^2$ obtained in the long wave approximation compared with the approximated solution for
		${\bar{\omega}} = \bar\mu$ and $\ell=0$. }
		\label{fig:3sub1}
	\end{subfigure}
	\begin{subfigure}[h!]{0.45\textwidth}
		\centering
		\includegraphics[width=\linewidth]{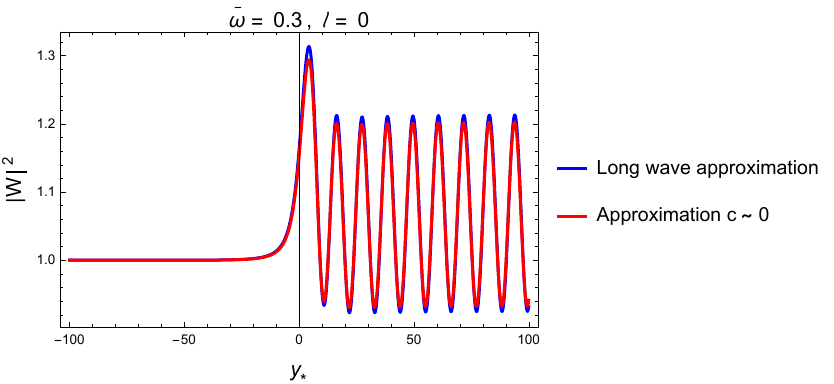}
		\caption{ $|W|^2$ obtained in the long wave approximation compared with the approximated solution for
		${\bar{\omega}} >\bar\mu$ and $\ell=0$.}\label{fig:3sub2}
	\end{subfigure}
	\begin{subfigure}[h!]{0.45\textwidth}
	\centering
	\includegraphics[width=\linewidth]{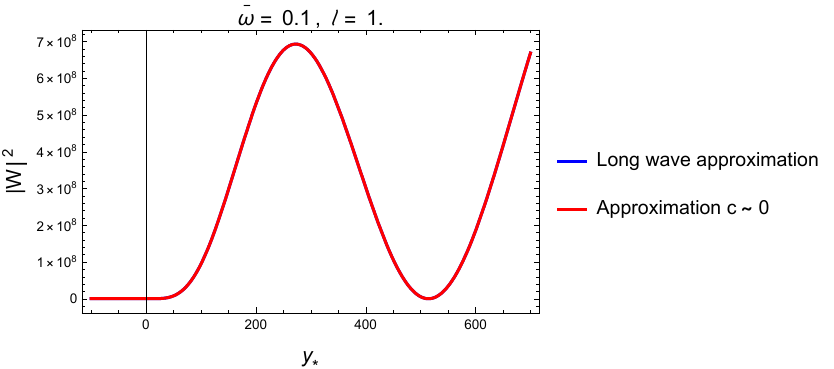}
	\caption{ $|W|^2$ obtained in the long wave approximation compared with the approximated solution for
		${\bar{\omega}} = \bar\mu$ and $\ell=1$.}\label{fig:3sub3}
\end{subfigure}
	\begin{subfigure}[h!]{0.45\textwidth}
	\centering
	\includegraphics[width=\linewidth]{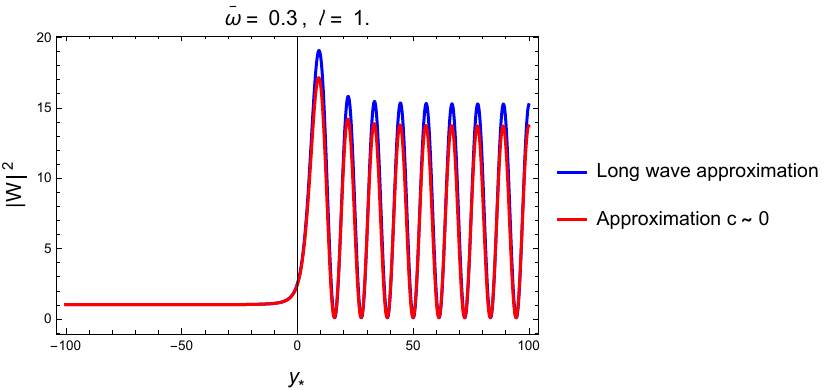}
	\caption{ $|W|^2$ obtained in the long wave approximation compared with the approximated solution for
		${\bar{\omega}} > \mu$ and $\ell=1$.}\label{fig:3sub4}
\end{subfigure}
	\caption{Compared amplitudes  $|W|^2$ for $\delta =  0.99$,  $\bar\mu=0.1$ and different $\ell$. In one case, we 
	use the long wave approximation, while the other corresponds to the solution calculated with the perturbative
	approach described in the Appendix \ref{sec:appendix1} }  \label{fig:FIG4}
\end{figure}

We can compare the solutions obtained by setting $c=0$ with those
coming from the general treatment in the Appendix \ref{sec:appendix1}.  The results are shown in  Figure \ref{fig:FIG4}.

The numerical solutions presented in figures {\ref{fig:FIG1}}, \ref{fig:FIG2}, \ref{fig:FIG3} and \ref{fig:FIG4} shows 
that: a) the source considered in the present work is relevant  in the region near the 
horizon, but in the limit $y_*\to \pm\infty$, the equation for the scalar field can be safely taken as (\ref{schtype}) with
the asymptotic form of the effective potential given in (\ref{asympV}) and sourceless; b) for numerical purposes, the choice $|y_*| \gtrsim 100$ is enough
to guarantee we are close enough to the horizon and infinity (depending on the sign of $y_*$); c) the case ${{\omega}}
\sim \mu$ can be treated using the long wave approximation, or the expressions for $\lambda_\ell$ obtained
in the Appendix \ref{sec:appendix1}.

In the following section, we will discuss the radiation pattern of the solutions analyzed here.

\section{Emission of Radiation}

Following \cite{detweiler}, the  emission of  axions radiation due to the source  (\ref{p1}),  per unit  frequency interval per
solid angle $d\Omega$ is  
\begin{equation}
\label{power}
\frac{d^2E_\ell}{d{\bar{\omega}}d\Omega} =\left( \frac{{S_\ell(\theta)}}{\sqrt{2\pi}}\right)^2\frac{1}{|2\,A^{(\infty)}(\bar\omega)|^2}
\left| \int_{-\infty}^\infty  W^{(\infty)} (y_*)\,\bar{A}_\ell  (y_*)\,dy_* \right|^2.
\end{equation}

The angular term $S_\ell^2$ depends on $c^2 =\omega^2-\mu^2$ but will omit this in the analysis since it represents 
a small contribution, as seen in the appendix, where this term is plotted as a function of the frequency.

The term $|A^{(\infty)}(\bar\omega)|^{-2}\equiv Q(\bar\omega)$ is obtained from the numerical solution of  (\ref{schtype})
with  initial conditions
  $$
  W(y_*) =e^{\imath\bar\omega y_*},\quad 
  W'(y_*)= \imath \bar\omega\,e^{\imath\bar\omega y_*},\quad (y_*\to-\infty),
  $$
and therefore
\begin{equation}
\label{A}
A^{(\infty)}(\bar\omega)=\lim_{y_*\to \infty}\left(
\frac{\sqrt{\bar\omega ^2-\bar\mu ^2} \,W(y_*)-\imath W'(y_*)}{2 \sqrt{\bar\omega^2-\bar\mu ^2}}
\right),
\end{equation}

Another interesting quantity to characterize the radiation emission is the fractional energy gain from
the monochromatic wave sent from infinity. In our case, this quantity is
\begin{equation}
\label{z}
Z=\left| \frac{C^{(\infty)}}{A^{(\infty)}} \right|^2-1,
\end{equation}
where $C^{(\infty)}$ is calculated in a similar way as $A^{(\infty)}$:
\begin{equation}
\label{C}
C^{(\infty)}(\bar\omega)=\lim_{y_*\to \infty}\left(
\frac{\sqrt{\bar\omega ^2-\bar\mu ^2} \,W(y_*)+\imath W'(y_*)}{2 \sqrt{\bar\omega^2-\bar\mu ^2}}
\right),
\end{equation}

\begin{figure}[h!]
	\centering
	\begin{subfigure}[h!]{0.45\textwidth}
		\centering
		\includegraphics[width=\linewidth]{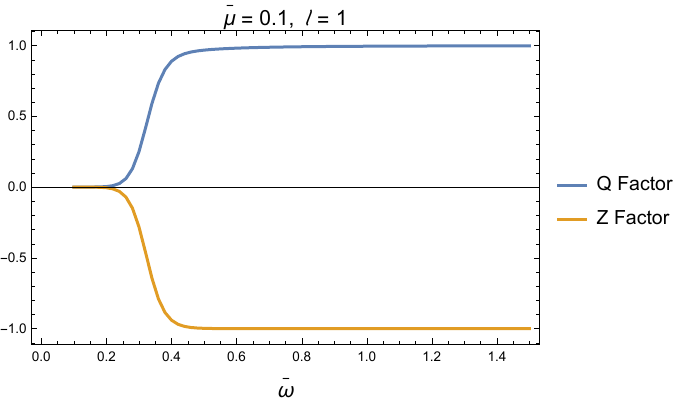}
		\caption{ Factors $Q$ and $Z$ for $\bar\mu=0.1$ and $\ell=1$. }
		\label{fig:5sub1}
	\end{subfigure}
	\begin{subfigure}[h!]{0.45\textwidth}
		\centering
		\includegraphics[width=\linewidth]{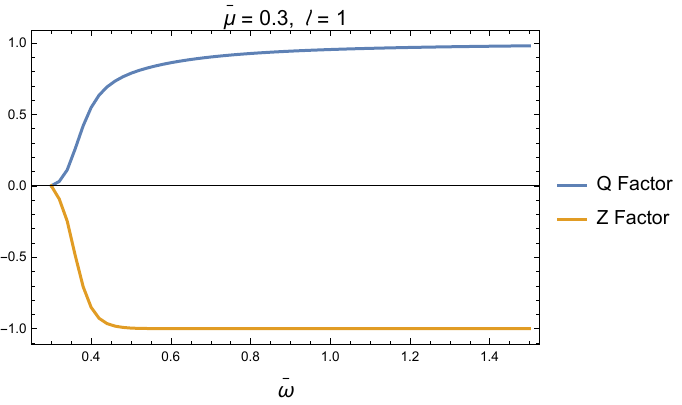}
		\caption{ Factors $Q$ and $Z$ for $\bar\mu=0.3$ and $\ell=1$}\label{fig:5sub2}
	\end{subfigure}
		\begin{subfigure}[h!]{0.45\textwidth}
		\centering
		\includegraphics[width=\linewidth]{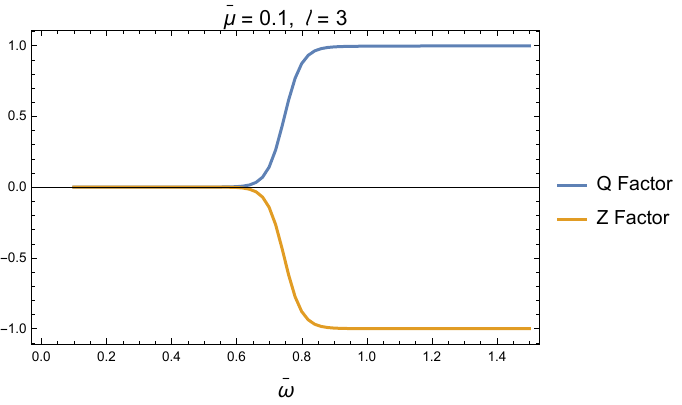}
		\caption {Factors $Q$ and $Z$ for $\bar\mu=0.1$ and $\ell=3$}\label{fig:5sub3}
	\end{subfigure}
		\begin{subfigure}[h!]{0.45\textwidth}
		\centering
		\includegraphics[width=\linewidth]{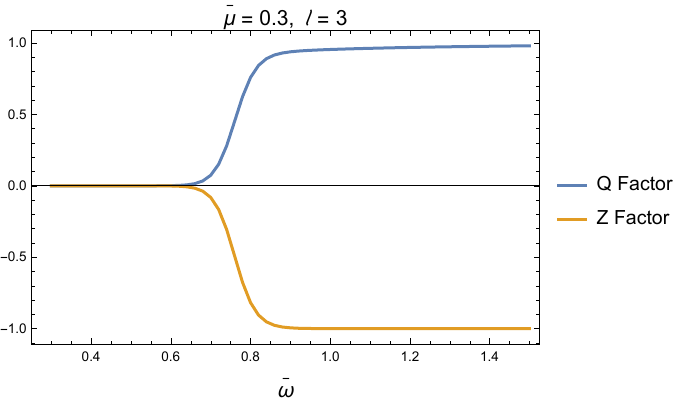}
		\caption{ Factors $Q$ and $Z$ for $\bar\mu=0.3$ and $\ell=3$.}\label{fig:5sub4}
	\end{subfigure}
		\caption{Factors $Q$ and $Z$ defined in the text for different values of $\ell$ and $\bar\mu$.
		 In all panels, $\delta=0.99$.}  \label{fig:FIG5}
\end{figure}

Figures \ref{fig:FIG5} show $Q$ and $Z$ for $\ell=0$ and $\ell=1$.  For the case $\ell=0$ (and for all even
values of $\ell$), the source term is zero; however, for even values of $\ell$, the source is relevant. 
Figures \ref{fig:5sub3} and \ref{fig:5sub4} show how different the factors $Z$ and $Q$ are when the source is considered. 

\begin{figure}[h!]
	\centering
	\begin{subfigure}[h!]{0.7\textwidth}
		\centering
		\includegraphics[width=\linewidth]{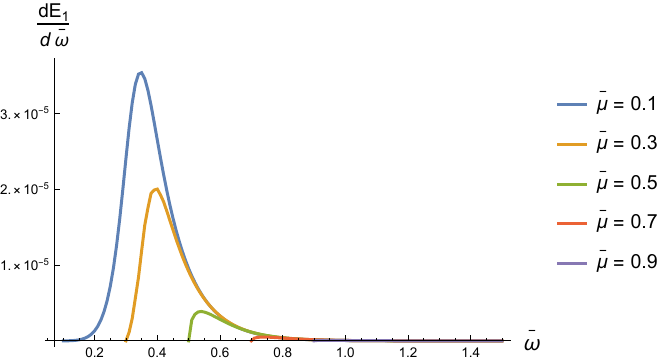}
		\caption{ Factors $Q$ and $Z$ for $\ell =0$. }
		\label{fig:6sub1}
	\end{subfigure}
		\caption{Radiated energy as function of $\bar\omega$ for different masses and $\ell=1,\delta=0.99$}  \label{fig:FIG6}
\end{figure}

With these results, we numerically calculate the total energy radiated to infinity up to the constant coming from 
the solid angle integration, that is 
\begin{equation}
\label{powerP}
\frac{d E}{d{\bar{\omega}}} =\frac{1}{|2\,A^{(\infty)}(\bar\omega)|^2}
\left| \int_{-\infty}^\infty  W^{(\infty)} (y_*)\,\bar{A}_\ell  (y_*)\,dy_* \right|^2,
\end{equation}
which is not zero only for odd values of $\ell$. In our case, this is $\ell=1,3$.  The results are plotted in 
Figure \ref{fig:FIG6}

\begin{figure}[h!]
	\centering
	\begin{subfigure}[h!]{0.7\textwidth}
		\centering
		\includegraphics[width=\linewidth]{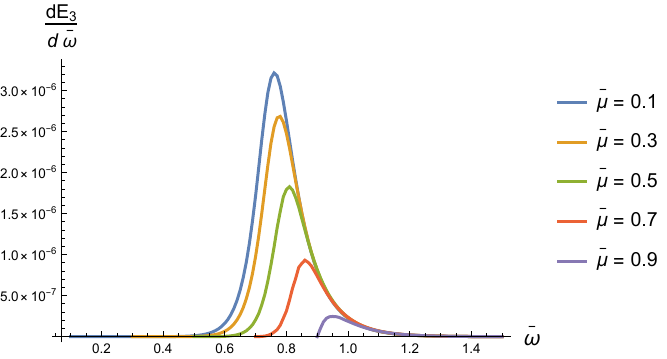}
		\caption{ Factors $Q$ and $Z$ for $\ell =0$. }
		\label{fig:6sub1}
	\end{subfigure}
		\caption{Radiated energy as function of $\bar\omega$ for different masses and $\ell=1,\delta=0.99$.}  \label{fig:FIG7}
\end{figure}

Let us comment on the results shown in this numerical analysis. According to Detweiler in  \cite{detweiler},  a sharp maximum in $Q$ signals a resonant frequency at which a black hole resonance occurs.  In our case, this resonance
does not happen, as seen in Figure \ref{fig:FIG5}. 

To understand this, first note that our initial condition (numerical integration condition) is $|W|^2 = |A^{(+)}|^2 = 1$ near the 
horizon ($y_* \to -\infty$ or, numerically, $y_* =-100$). On the other hand, functions $Z$ and $Q$ start from zero at the initial frequency $\bar\omega = \bar\mu$, then $Z$ increase while $Q$  decrease, which happens during a frequency interval, let's say $\Delta \bar{\omega}$. By denoting the half of such interval as $\bar{\omega}_c$, the function $Z$
behaves as follow
\begin{equation}
\label{qbehave}
Z = \left\{
\begin{array}{lcl}
0   &   & \bar{\omega} \lessapprox \bar{\omega}_c  - \Delta\bar{\omega},
\\
-1   &    &\bar{\omega} \gtrapprox \bar{\omega}_c +\Delta\bar\omega,
\end{array}  \right.
\end{equation}
and a similar expression for $Q$, changing the last line to $1$.

From the definition of $Q$ and $Z$, previous behavior is understood due to the following. For $\bar{\omega} \lessapprox \bar{\omega}_c - \Delta\bar\omega$, the denominator in (\ref{A})  produces a divergence that is the responsible for 
$Q\to 0$. Instead, for $Z$, such divergence is not present since $Z$ depends on the ratio $ C^{(\infty)}/A^{(\infty)}$ 
(see the definition of $C^{(\infty)}$ in (\ref{C}) and then for this frequency range, $ |C^{(\infty)}|^2\sim |A^{(\infty)}|^2$

Instead, for $\bar{\omega} \gtrapprox \bar{\omega}_c +\Delta\bar\omega$, the condition  $ |A^{(\infty)}|^2 \sim 1$ and 
$|C^{(\infty)}|^2\sim 0$ is consistent with the behavior of $Z$ and $Q$ factors.

Since the  $Z$ factor is the fractional energy gain of a wave of frequency 
$k=\sqrt{\bar\omega^2-\bar\mu^2}$ sent from the infinity that is scattered from the BH. In the zone where $Z\sim 0$, 
part of the incoming wave is scattered (indeed, in this frequency range $|C^{(\infty)}|\sim |A^{(\infty)}|$, while in the
region in which $Z\to-1$, no scattered wave is present, indicating a complete absorption of the signal.

Therefore, the energy radiated should be centered in the transition zone, the $\Delta\omega$ region.  Figures \ref{fig:FIG6}  and \ref{fig:FIG7} precisely show this behavior.

\section{Conclusions}

This research paper explores the movement of gravitational axions in a Kerr black hole background 
using analytical and numerical methods. One interesting finding is that resonance occurs 
when $\omega \gtrsim \mu$, similar to the Detweiler-resonance discussed in \cite{detweiler}. 
However, the Detweiler-resonance is related to a massless scalar. This research concludes that 
this resonance always occurs if $\ell$ is odd and a Pontryaguin-like source is present.

Another important observation is that Figures 6 and 7 show that the spectral maxima shifts to
the right, and the radiated power decreases with ${\bar \omega}$. Additionally, the 
maximum becomes significantly smaller when $\ell$ grows. However, it is interesting to note
qualitatively that the curves can be reasonably approximated as Gaussian for small $\bar\mu$.

For $\bar\mu\gtrsim 0.3$, the Gaussian starts to be asymmetric with a deviation to the right. We found 
that the function
\begin{equation}
\label{skew}
f(\bar{\omega}) = a\, e^{-\frac{(x-\xi)^2}{2\,\sigma^2}}
\left(
1 + \mbox{\small{erf}} \left(\frac{\alpha\,( x - \xi )}{\sqrt{2}\,\sigma}\right),
\right)
\end{equation}
is well-fitted to the curves of the radiated power.  Here, $ \mbox{\small{erf}} (x) $ is the error function
$$
 \mbox{\small{erf}} (x) = \frac{2}{\sqrt\pi}\int_0^x e^{-t^2}\,dt.
$$

The function $f$ in (\ref{skew}) is proportional to the probability density of a {\it skew-normal} distribution,
with proportionality constant $a$ and for this distribution, it is known that the mean value is
$$
\mbox{Mean} = \mu +\frac{\sqrt{\frac{2}{\pi }} \alpha  \sigma }{\sqrt{1+\alpha ^2}},
$$
and the variance
$$
\mbox{Var} = \sigma^2\left(1-\frac{2\alpha^2}{\pi(1+\alpha^2)}\right).
$$  
So, an estimation of the mean width of the curves in  Figures \ref{fig:FIG6} and \ref{fig:FIG7} is 
given by $2\sqrt{\mbox{Var} }$. That is 
$$
\Delta\bar\omega \approx 2\sigma \sqrt{1-\frac{2\alpha^2}{\pi(1+\alpha^2)}}.
$$

The following table resumes these results
\begin{table}
\caption*{The mean value and variance  for different $\ell$ and $\bar\mu$ from Figures \ref{fig:FIG6} and \ref{fig:FIG7} }
\begin{center}
\begin{tabular}{ |c|c|c|c| } 
\hline
$\ell$& $\bar\mu$ & $\langle\bar{\omega}\rangle$ &$\Delta\bar\omega$ \\
\hline
\multirow{5}{3em}{~~~~1}
& ~~0.1~~ & ~~0.38~~ &~~0.16~~  \\ 
& ~~0.3 ~~&  ~~0.44~~ &~~0.16~~  \\ 
& ~0.5~~&  ~~0.61~~ &~~0.15~~  \\ 
& ~~0.7 ~~&  ~~0.81~~ &~~0.15~~  \\ 
& ~~0.9~~ &  ~~1.01~~ &~~0.15~~  \\ 
\hline
\multirow{3}{3em}{~~~~3} 
& ~~0.1~~ & ~~0.78~~ &~~0.15~~  \\ 
& ~~0.3 ~~&  ~~0.80~~ &~~0.15~~  \\ 
& ~0.5~~&  ~~0.84~~ &~~0.14~~  \\ 
& ~~0.7 ~~&  ~~0.90~~ &~~0.14~~  \\ 
\hline
\end{tabular}
\end{center}
\end{table}

Estimating the mean lifetime of these resonances as (dimensionful  quantities)
$$
\tau \approx (\Delta \omega)^{-1} = \frac{M}{\Delta\bar\omega} \approx  5\times10^{38} 
\left(\frac{M}{M_\odot} \right)
~~~[{\mbox{s}}]
$$

Then, for example,  for primordial black holes with the masses ranging from  Planck mass ($\sim 10^{-37} M_\odot$) to 
 masses of order $10^3 M_\odot$ will cause pulses with a mean lifetime from  $10^{-1}$ [s] up to $10^{41}$ [s]$\sim 10^{34}$ years, that is (in the last case) a mean lifetime well beyond the universes's age. 

It's worth noting that the resonance for gravitational axions discussed in this 
paper is distinct from the one studied in Detweiler's work. In Detweiler's case \cite{detweiler}, 
the resonance is observed as a divergence of the factor $Q({\bar \omega})$, 
whereas in our case, $Q({\bar \omega})$ approaches 1 as ${\bar \omega}$
 tends to infinity.
 
{{In conclusion, observing very sharp resonances in radiation patterns is possible depending on the black holes' mass. For instance, if the black holes have masses between 100 and 1000   $M\textsubscript{\(\odot\)}$, the lifetime of these resonances falls within the range of observability in LIGO \cite{soda}.}}
 
\noindent 

\section*{Acknowledgements}

  J.G.  thanks  Prof. Christophe Grojean for the pleasant
  hospitality in DESY (Hamburg); he also thanks Thomas Biek\"otter, Mathias Pierre, for the pleasant discussions at the DESY lunch and also mainly to Andreas Ringwald and Pierre Sikivie for sharing his knowledge of axions with him.  The Alexander von Humboldt Foundation financed J.G.'s work.  This research was supported by  Fondecyt 1221463 (J.G.) 
and DICYT-USACH  042231MF (F.M.).

\appendix

\section{The angular equation}
\label{sec:appendix1}

In this appendix, we review the solution of (\ref{oblate}) following
the analysis by S. Teukolsky in (\cite{teukolsky}).  First, define as
usual $x = \cos\theta$, so that the angular equation is now 
\begin{equation}
  \label{angular1}
  \frac{d}{dx} \left[(1-x^2)\frac{d {S}}{dx} \right] + (\lambda +
  c^2x^2){S}=0.
\end{equation}
In reference (\cite{teukolsky}), a method
for treating the case of any spin $s$ and third component of angular
momentum, $m$,   is discussed. However, we restrict here to the case of
interest for us, that is, $m=0$ and $s=0$,  which is just previous
equation. 

The idea is to treat the $c^2 x^2$ term as a perturbation (not necessarily infinitesimal). 
The order zero operator is  
\begin{equation}
	\frac{d}{dx}\left[(1-x^2)\frac{d {S}}{dx}\right] = -\lambda {S},
\end{equation}
with the known (normalized) solution
\begin{equation}
  \label{szero}
	{S}_\ell(x) = \sqrt{\frac{2\ell+1}{2}} P_\ell(x),
\end{equation}
with    $\lambda =\ell(\ell+1)$ ( $\ell =0,1,2\cdots$), and
$P_\ell(x)$ the Legendre's polynomial of degree $\ell$.

The continuation method \cite{wasserstrom} for calculating
eigenfunction and eigenvalues in (\ref{angular1}) considers $c$ as a
parameter, and then the equation under study 
is 
\begin{equation}
	\label{angular2}
	\left[(1-x^2) {{S}'}_\ell\right]'+
	(\lambda_\ell(c) + c^2 x^2){S}_\ell=0,
\end{equation}
with $'$ denoting derivatives respect to $x$ and ${S}_\ell = S_\ell(c,x)$.

By taking the  derivative with respect to $c$ in the previous equation (denoted by a dot in what
follows), one obtains 
\begin{equation}
	\label{angularc}
	\left[(1-x^2)\dot{{S}'_\ell}\right]' + (\dot{\lambda}_\ell 
	+ 2 c x^2){S}_\ell + ( \lambda_\ell + c^2 x^2)\dot{S}_\ell =0. 
\end{equation}

From here, it is possible to find a set of  first-order differential equations 
for $\lambda_\ell(c)$ and ${S}_\ell$. Indeed, multiplying the last equation by  
${S}_\ell$ followed by an $x$ integration one gets 
\begin{equation}
  \label{angularint}
  \int dx {S}_\ell	\left[(1-x^2)\dot{{S}'_\ell}\right]'  +
  \int dx {S}_\ell(\dot{\lambda}_\ell + 2 c x^2){S}_\ell  +
  \int dx {S}_\ell(\lambda_\ell + c^2 x^2)\dot{{S}_\ell} =0.
\end{equation}

The first term can be integrated by parts twice giving  (boundary terms cancel) 
\begin{eqnarray}
  \int dx~ {S}_\ell \left[(1-x^2)\dot{{S}'_\ell}\right]' &=&
  \int dx~ \dot{{S}}_\ell \left[(1-x^2){{S}'_\ell}\right]',
  \nonumber
  \\
  &=-&\int dx~ \dot{{S}}_\ell (\lambda_\ell + c^2 x^2){S}_\ell.
\end{eqnarray}
Then, the first and second terms cancel in (\ref{angularint}). Finally, 
\begin{equation}
	\label{lambdac}
	\dot{\lambda}_\ell =-~\frac{2c}{|{S}_\ell|^2} \int dx~{S}_\ell~x^2~{S}_\ell,
\end{equation}
with $\displaystyle{|{S}_\ell|^2 =\int{S}_\ell{S}_\ell dx}$.

We repeat  the calculation, but multiplying now by ${S}_{\bar{\ell}}$ with 
$ \bar{\ell}\neq\ell$, and  performing the integral to obtain
\begin{equation}
	\int 	\dot{{S}}_\ell{S}_{\bar{\ell}}~ dx =-2c	\int 
	\frac{ {{S}}_\ell \, x^2\,{S}_{\bar{\ell}}}{\lambda_\ell - \lambda_{\bar{\ell}}}dx.
\end{equation}
Finally, with the help of  completeness relation, one obtains
\begin{equation}
	\label{zeq}
	\dot{{S}}_\ell = -2 c \sum_{\bar{\ell}\neq\ell}
	 \frac{{S}_{\bar{\ell}}(x)}{\lambda_\ell - \lambda_{\bar{\ell}}}
	 	\int  {{S}}_\ell(x') \, x'^2\,{S}_{\bar{\ell}}(x')\,dx'.
\end{equation} 

Equations (\ref{lambdac}) and (\ref{zeq}) are a system of  differential equations 
to be solved (numerically) to determine the eigenvalues
$\lambda_\ell(c)$  and eigenfunctions $S_\ell(c,x)$ in equation (\ref{angular1}).

In this perturbative approach, the solution of the problem at zero
order is given by (\ref{szero}), then  we look for solutions of
(\ref{lambdac}) and (\ref{zeq})  with the form
\begin{equation}
	\label{Zexpan}
	{S}_\ell(c,x) = \sum_{\ell',\ell''} B_{\ell,\ell'}(c)
        \sqrt{\frac{2\ell'+1}{2}} P_{\ell'}(x),
\end{equation}
which, once replaced in the equations, give rise to  
\begin{eqnarray}
        \label{finallambda}
	\frac{d\lambda_\ell}{d c} &=& - \frac{2c}{|B_{\ell,\ell}|^2}
        \sum_{\ell',\ell''}  B_{\ell, \ell'} B_{\ell, \ell''}
        \langle \ell'|\ell''\rangle 
        \\
        \label{finala}
	\frac{dA_{\ell,\ell'}}{d c} &=& -2c
        \sum_{\substack{\bar{\ell}\neq\ell \\ L',L'' }}
        \frac{B_{\bar{\ell},\ell'}        }{
          \lambda_\ell-\lambda_{\bar{\ell}}} B_{\ell,L'} \langle
        L'|L''\rangle B_{\bar{\ell},L''}, 
\end{eqnarray}
with
$$
\langle m | n\rangle =\frac{[(2m+1)(2n+1)]^{1/2}}{2} \int_{-1}^1 P_m(x)\,x^2\,P_n(x)dx,
\quad
|A_{\ell,\ell}|^2 = \sum_{\ell'}(A_{\ell,\ell'})^2.
$$
Equations must be solved with the following initial conditions
\begin{equation}
  \lambda_\ell(0) =\ell(\ell+1),
  \quad
  B_{\ell,\ell'}(0) = \delta_{\ell,\ell'}.
\end{equation}

Expressions (\ref{finallambda}) and (\ref{finala}) are those in
(\cite{teukolsky})\footnote{In the 
Teukolsky's approach, the normalization  $| {S}_\ell |^2 = 1$ is
assumed. }, specified in our case for the  massive  scalar field and
the third component of angular momentum equals zero.

The case $c^2 < 0$ can be treated similarly, and the only effect
is a change of signs in the RHS of  (\ref{finallambda}) and
(\ref{finala}). However, this case is not interesting since it
produces divergent solutions for $r\gg r_+$.

Our analysis focuses on the cases  $c \geq  0$. The numerical
solutions are obtained by summing  up to $\ell' = 10$ in
(\ref{Zexpan}) and subsequent expressions. The solutions are 
fitted to a polynomial  function, and we found that the best fit (for 
$0 < c < 5$) is obtained for order five or higher polynomials. 

The results for eigenvalues $\lambda_{\ell}$ with  $\ell=0,1,2,3$ are 
the following
\begin{eqnarray} 
  \lambda_0 (c) &=&
  -0.0298 \,c - 0.297 \,c^2 - 0.00746 \,c^3 - 0.0372 \,c^4 + 0.00352 \,c^5
  \\
  \lambda_1 (c) &=&
  2 - 0.00112\, c -0.6\, c^2 +0.00294\, c^3   -0.00964\, c^4+   0.000834\, c^5
  \\
  \lambda_2 (c) &=&
  6 + 0.00412\, c - 0.5350 \, c^2 + 0.00998 \, c^3 - 0.00844 \, c^4 + 0.000553 \, c^5
  \\
  \lambda_3 (c) &=&
  12 + 0.0018\,c -0.513\,c^2 -0.00176\, c^3 + 0.00572 \, c^4 - 0.000794\, c^5
\end{eqnarray}

The coefficients $B$ in (\ref{Zexpan}), on the other hand, are the following
\begin{eqnarray} 
  B_{1,1} (c) &=&
     1 - 0.000774 \, c + 0.00169 \, c^2 - 0.00104 \, c^3 - 0.000193 \,
     c^4 + 0.0000372 \, c^5
  \\
  B_{1,3} (c) &=&
     0.00156 \, c  + 0.022 \, c ^2 + 0.00366   \, c ^3 - 0.00148 \, c
     ^4 + 0.000105 \, c ^5
 \\
     B_{3,1} (c) &=&
 -0.00144  \, c - 0.0223  \, c^2 - 0.00339  \, c^3 + 0.00139  \, c^4 -
 0.0000965  \, c^5
  \\
  B_{3,3} (c) &=&
  1 - 0.000666  \, c + 0.00143  \, c^2 - 0.000842  \, c^3 - 0.000355
  \, c^4 + 0.0000446  \, c^5
\end{eqnarray}

These coefficients are the non-zero ones, which are relevant to our approximation.  For example, 
\begin{eqnarray}
{S}_1  &=& \sqrt{\frac{1}2}\,B_{1,1}(c)\,P_1(\cos\theta) +
\sqrt{\frac{5}2}\,B_{1,3}(c)\,P_3(\cos\theta)+\cdots, 
\nonumber
\\
{S}_3  &=& \sqrt{\frac{3}2}\,B_{3,1}(c)\,P_1(\cos\theta) +
\sqrt{\frac{7}2}\,B_{3,3}(c)\,P_3(\cos\theta)+\cdots.
\nonumber
\end{eqnarray}

Finally, note that for the radiation emission, the quantities of interest are $|S_\ell|^2$, which, once integrated
into the solid angle, will be $1$.  However, our approach has an explicit dependence on $c$
\begin{eqnarray}
\int_{-1}^1 |{S}_1|^2 \,d(\cos\theta) &=& |\,B_{1,1}(c)|^2 + |B_{1,3}(c)|^2+\cdots, 
\nonumber
\\
\int_{-1}^1|{S}_3 |^2\,d(\cos\theta) &=&  |B_{3,1}(c)|^2 + |B_{3,3}(c)|^2 +\cdots.
\nonumber
\end{eqnarray}

\begin{figure}[h!]
	\centering
		\includegraphics[width=\linewidth]{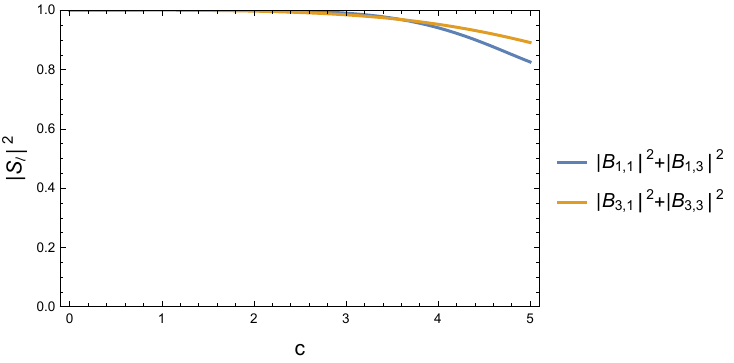}
			\caption{Contribution of $c = \delta\sqrt{\bar\omega^2 -\bar\mu^2}$ to the normalization of 
			$S_\ell$ functions}  \label{figapp:FIG1}
\end{figure}

Figure \ref{figapp:FIG1} shows that even if there is such a dependence, these integrals can be approximated to 1
in our numerical analysis.


\end{document}